\documentclass{elsart}

\usepackage{graphicx}

\usepackage{amssymb}

\begin{document}

\begin{frontmatter}


\title{Energy gap of
ferromagnet-superconductor bilayers}
\author{Klaus Halterman}
\address{Sensor and Signal Sciences Division, Naval Air Warfare Center,
China Lake, California 93355\thanksref{thanks}}
\ead{klaus.halterman@navy.mil}
\author{Oriol T. Valls}
\address{School of Physics and Astronomy and Minnesota Supercomputer
Institute, Minneapolis, Minnesota 55455}
\ead{otvalls@umn.edu}
\thanks[thanks]{Supported in part by a grant of HPC time
 from the DoD HPC 
Center, NAVO.}





\begin{abstract}
The excitation spectrum of 
clean
ferromagnet-superconductor
bilayers is calculated  within the 
framework of the self-consistent
Bogoliubov-de Gennes theory. Because of the proximity
effect,  
the superconductor  induces a gap in the ferromagnet spectrum,
for thin ferromagnetic layers.
The effect depends  strongly on the  exchange 
field in the ferromagnet.
We find that as the thickness of the  ferromagnetic layer increases,
the gap disappears, and that its destruction  arises from
those quasiparticle excitations with wavevectors
mainly along the interface. 
We  discuss the influence
that the interface quality and Fermi energy mismatch 
between the ferromagnet and superconductor have on the
calculated energy gap. We also evaluate the density of states in
the ferromagnet, and we find it in all cases consistent with the gap
results.
\end{abstract}

\end{frontmatter}



\section{Introduction}
Through the
mechanism of Andreev reflection\cite{andreev}, a
normal (non-superconducting)
material in contact with a superconductor acquires
superconducting phase coherence between its particle
and hole wave functions, even though the pairing
interactions vanish in this normal material. 
The associated superconducting correlations give rise to a finite value
of the pair amplitude,
$F({\bf r}) = \langle \hat{\psi}_{\uparrow} ({\bf r})  \hat{\psi}_{\downarrow} 
({\bf r}) \rangle$. This is known as the proximity
effect. 

The proximity effect occurs also, in a modified way, if the
non-superconductor is a ferromagnet. In this case, $F({\bf r})$ exhibits
spatial oscillations.
On the other hand, the pair potential itself, $\Delta ({\bf r})$ vanishes in
the ferromagnet in the absence of  attractive coupling. 
For conventional isotropic superconductors, $\Delta$
is spatially independent, and
corresponds to the 
minimum excitation energy in the spectrum,
or the
energy gap, $E_g$.
For  ferromagnet-superconductor
systems, as in other inhomogeneous cases, $ \Delta({\bf r})$ 
depends on position,
while the excitation gap for the whole system, $E_g$,
obviously  does not. 
The proximity effect can therefore result in 
$E_g$ and $\Delta$ possessing a rather
nontrivial relationship. In this paper we
study this question for ferromagnet-superconductor (F/S) bilayers.

Several works involving
nonmagnetic normal metal-superconductor 
heterostructures have discussed in
various contexts, the presence or absence of
an energy gap in the excitation 
spectrum.\cite{zhou,melson,belzig,ostrovsky,ivanov}
For a normal metal layer of width greater than the coherence length in the
superconductor, $\xi_0$,
but smaller than the phase coherence length $L_\phi$,
the appearance of a minigap 
of order of the Thouless energy has been reported,
while if the normal layer is smaller than $\xi_0$, the induced gap is of 
order $\Delta$.
It was found that while
interface roughness plays a role,
the minigap persistence depends
chiefly upon interface quality.\cite{pilgram}
These conclusions  draw upon the results
of calculations carried out within the
quasiclassical formalism, which is applicable when
certain conditions are satisfied.
The main results are modified
if the normal metal region is modeled as a 
non-Fermi liquid, where
increased electron correlations may 
inhibit or completely destroy the minigap altogether.\cite{branislav}

For  F/S structures, the magnetic exchange field
in the ferromagnet results in quasiparticle spin splitting, and 
consequently the  
pair amplitude rapidly decays and oscillates over a characteristic length scale $\xi_F$,
in which electron-hole pairs remain coherent.
This modulation in phase of the pair amplitude has been shown to induce
corresponding oscillations in the local density of states 
(DOS).\cite{buzdin,proximity,bve,zareyan}
Impurities in the magnet were shown
to play a significant role in 
the determination of
thermodynamic quantities.\cite{buzdin,bve}
For ballistic F/S structures,
the tunneling DOS for clean ferromagnet layers has
been calculated for a wide range of exchange fields,\cite{zareyan}
and the local DOS has been found to exhibit a gapless state
for all positions in the magnet.\cite{proximity,xing}
Configurations 
consisting of weak or 
thin ferromagnetic layers where  $d_F/\xi_F\ll 1$,
may lead to significantly different  results,\cite{vecino} as
we shall see below.
The finite geometry can 
result in the quasiparticle amplitudes 
undergoing coherent Andreev and normal reflection
(at the ferromagnet-insulator interface) in such
a way that the minimum value
of the excitation spectrum is nonzero
due to the relatively slow
characteristic decay of the pair amplitude. Thus, a gap develops.
Associated with this,
when the dimensions of the ferromagnet are reduced
to a scale comparable to $\xi_F$,
the greater confinement of the pairing correlations
in the ferromagnet
restricts the pair amplitude so that it no longer changes sign, and the
DOS does not exhibit the corresponding oscillations.

It is generally accepted that
the quasiclassical and dirty limit equations are
satisfactory for studying 
many aspects of
inhomogenous superconductivity. 
If however, the
ferromagnet has dimensions of order of or smaller than the elastic 
mean free path $\ell$,
so that quasiparticles undergo very few scattering events,
the problem is more suitably solved within the ballistic regime. Furthermore, 
the inherent geometrical
effects associated with ferromagnet layers only a few atomic spacings 
thick require
a full microscopic theory that does not coarse-grain over
length scales of order of the Fermi wavelength $\lambda_F$. Another constraint
imposed upon the quasiclassical approximation is exhibited
by the Andreev\cite{andreev,geof} equations,
whereby values of the quasiparticle momentum parallel to
the interface, and
comparable to the Fermi momentum, are crudely approximated.
We will see below that in fact these quasiparticle states play a prominent role
in the determination of the gap threshold for F/S systems.

Thus, we  examine here the 
electronic spectrum of F/S bilayers and investigate
under what conditions an energy gap can
exist in such structures.
The problem will be solved
within the framework of the
{\it self-consistent} solutions
of the Bogoliubov de-Gennes\cite{bdg} (BdG) equations.
We will  calculate self-consistently the
DOS, and
excitation spectra of a thin ferromagnet film adjoining a 
bulk superconductor.
This microscopic approach 
complements existing quasiclassical theories and serves to round the 
current knowledge. 

In the next section, we introduce the
method and geometry used in our model. In Sec.\,\ref{results}, 
the influence of the interface transparency, the magnitude of the exchange field,
and the mismatch in Fermi levels on the energy gap and local DOS, will be presented.

\section{Geometry and model \label{method}} 

We consider a heterojunction of total length $d$ in the
$z$-direction, consisting 
of a ferromagnet of thickness $d_F$ and a superconductor of thickness
$d_S$. The planar interface is at $z=d_F$, 
and the free surfaces at $z=0$ and $z=d$ are specularly reflecting.
Upon taking into account the translational invariance in the 
semi-infinite $x-y$ plane, one 
can immediately write down the  BdG equations\cite{bdg,proximity}
for the spin-up 
and spin-down quasiparticle and quasihole wave functions 
$(u_n^\uparrow,v_n^\downarrow)$,
\begin{eqnarray}\label{bogo}
\Bigl[-\frac{1}{2m}\frac{\partial^2}{\partial z^2} 
+\varepsilon_{\perp} 
+W(z)-E_F(z)\Bigr] 
u^{\uparrow}_n(z) 
+
\Delta(z) v^\downarrow_n(z)& = &\epsilon_n u^{\uparrow}_n(z),
\label{bogo1} 
 \\ 
-\Bigl[ -\frac{1}{2m}\frac{\partial^2}{\partial z^2} 
+\varepsilon_{\perp} +W(z)
-E_F(z)\Bigr] 
v^\downarrow_n(z) 
+ \Delta(z) u^\uparrow_n(z)& = & 
\epsilon_n v^\downarrow_n(z), \label{bogo2} 
\end{eqnarray}
where 
$\varepsilon_{\perp}$ is the transverse kinetic 
energy, $\epsilon_n$ are the 
quasiparticle energy eigenvalues,
$h_0(z)=h_0 \Theta(z-d_F)$ is the magnetic 
exchange energy, and $\Delta(z)$ is the pair potential.
Scattering at the interface is modeled by
the potential $W(z)=W \delta(z-d_F)$, where $W$ is the barrier
strength parameter. We take the quantity
$E_F(z)$ to
equal  $E_{FM}$ in the magnetic
side, 
so that $E_{F\uparrow}\equiv E_{FM}+h_0$, and $E_{F\downarrow}\equiv E_{FM}-h_0$,
while  in the superconducting side,
$E_F(z)\equiv E_{F}$.
From the symmetry of the problem, 
the solutions for the other set of
wavefunctions  
$(u_{n}^{\downarrow},v_{n}^{\uparrow})$  
are  easily obtained  
from those of Eqns.~(\ref{bogo1},\ref{bogo2})
by allowing for both 
positive and negative energies.
The BdG equations are supplemented by the self
consistency condition for the pair potential,
\begin{equation}  
\label{del2} 
\Delta(z) =\frac{g(z)}{2} 
\sum_{\epsilon_n\leq \omega_D} \left[
u_n^\uparrow(z)v^\downarrow_n (z)+
u_n^\downarrow(z)v^\uparrow_n (z)\right]\tanh(\epsilon_n/2T), 
\end{equation} 
where $T$ is the temperature, 
$g(z)$ is the effective  coupling constant, describing 
the electron-electron interaction, and $\omega_D$
is the Debye energy.

\section{Results \label{results}} 
In the following, we consider 
a thin ferromagnet layer
adjacent to a
bulk limit $(d_S\gg\xi_0)$ superconductor.
To focus on the parameter range which may yield
a gap in the energy spectrum,
the thickness of the
ferromagnet considered will not exceed the
coherence length in the superconductor, i.e.,
$d_F<\xi_0$.
We introduce the dimensionless parameter $Z$ to characterize
the barrier strength, with $Z\equiv m W/k_F$, and the measure
of mismatch between the Fermi levels is given by the 
ratio $\Lambda\equiv E_{FM}/E_F$.
The requirement of self-consistency in the problem
excludes the possibility of an analytical solution.
The BdG equations (\ref{bogo1},\ref{bogo2}) are therefore solved numerically 
using the approach\cite{proximity,klaus} outlined in the Appendix.
We assume the low temperature regime (we take $T/T_c=0.02$), and
a superconducting coherence length $\xi_0$ of $k_F \xi_0 =50$, where
$k_F$ is the Fermi wavevector of the superconductor. 

The energy gap, $E_g$, is the minimum binding energy of a Cooper pair, 
and its existence will influence
various thermodynamic measurements, including 
heat capacity and thermal conductivity. For other configurations,
such as 
long superconductor-normal metal-superconductor junctions,
the gap in the metal and the bulk gap in the superconductor can further
inhibit thermal transport.
\begin{figure}
\includegraphics{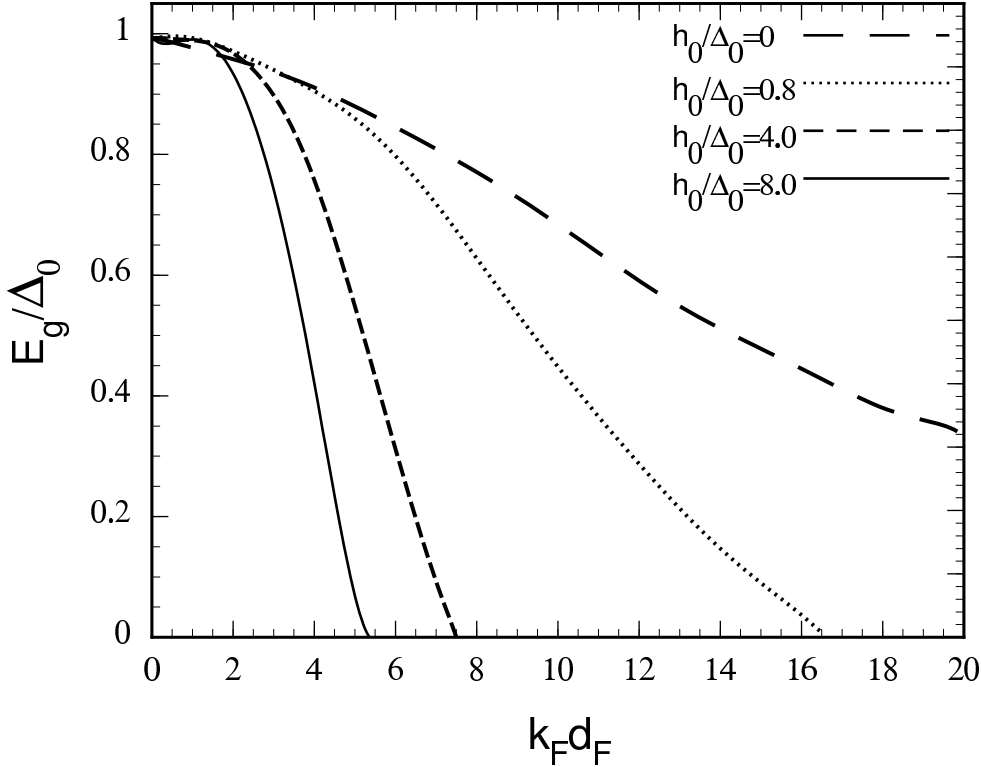}
 
\caption{Variation of the 
normalized energy gap in the
ferromagnet with the
dimensionless
ferromagnet thickness $k_F d_F$.
The coherence length is given by $k_F \xi_0 =50$.
The different curves correspond to 
the four values of the magnetic exchange energy
labeled in the figure. \label{gap}}
 
\end{figure}
The process of finding $E_g$ is computationally demanding,
as it involves calculating self-consistently
the entire eigenvalue spectrum and then finding its
minimum, for each value of $d_F$. 
In Fig.\ref{gap} we illustrate our results for the dependence of 
$E_g$ on the thickness of the ferromagnet layer.
These results are for $Z=0$ and $\Lambda=1$. 
Results for four different exchange fields in the ferromagnet are presented.
The top dashed curve corresponds to $h_0/\Delta_0=0$, while
the subsequent lower curves are for finite values of the exchange field
as labeled in the figure. 
All curves start at $E_g=\Delta_0$ for $d_F=0$, 
corresponding to the expected result for a single superconductor
in the bulk limit. As the ferromagnet layer grows,
$E_g$
decays  monotonically towards a gapless superconducting
state, at a rate that
depends strongly on
the strength of $h_0$. In the limit of zero exchange field (normal metal),
the gap ceases to exist when $ d_F\approx \xi_0$. This thickness at
which $E_g$ is destroyed is much smaller than the 
proximity length characterizing the usual
very slow decay of the pair amplitude in the metal.\cite{falk}
This trend continues 
for finite $h_0$, with
$E_g$ vanishing much more rapidly than the 
characteristic decay length of the pair amplitude
in the magnet. To illustrate this, recall\cite{klaus,demler}
the approximate expression
for the characteristic length of decay of the pair amplitude
in the ferromagnet:
$\xi_F \approx (k_{F\uparrow}-k_{F\downarrow})^{-1}$,
which can be rewritten as,
\begin{equation} \label{length}
k_{F M} \xi_F \approx  \frac{\pi}{2} \Lambda^{3/2}\left(\frac{\Delta_0}{h_0}\right) k_F \xi_0.
\end{equation}
Thus for example, the $h_0/\Delta_0=4.0$ curve in Fig.~\ref{gap} shows the 
energy gap
vanishing at $k_{FM} d_F  \approx 7.5$, while the pair amplitude decays 
over the larger length
scale $k_{FM} \xi_F \approx  20$. The discrepancy is even larger
if one recalls that the expected decay length might be thought to
be a factor of $2 \pi$ larger.
For 
the largest field shown ($h_0/\Delta_0=8$),
the superconducting correlations 
are further inhibited, failing to generate a
gap in the energy spectrum except for ferromagnet
nanostructures satisfying
$k_F d_F\lesssim 5$.

\begin{figure}
\includegraphics{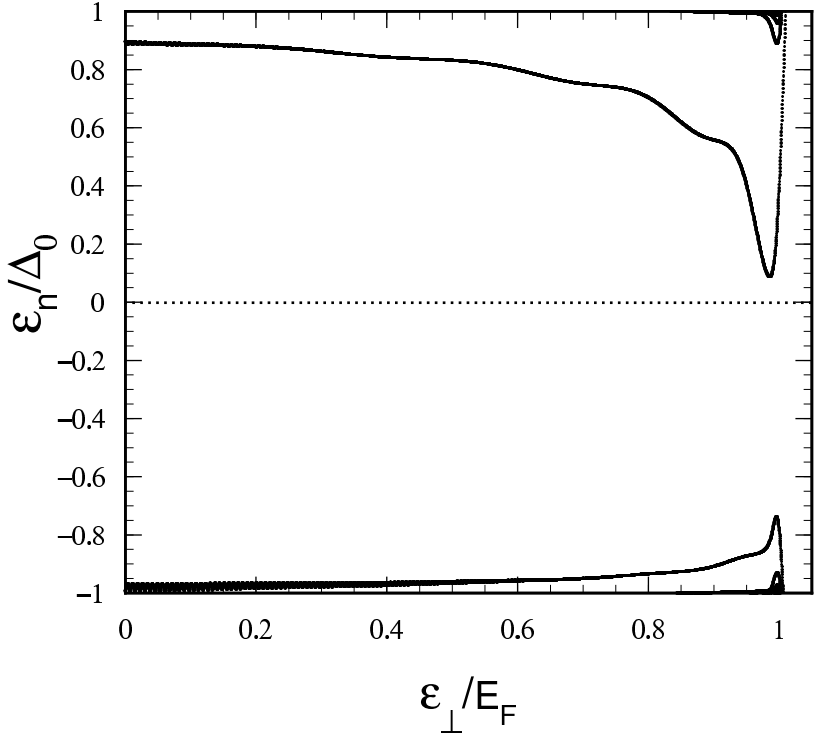}
 
\caption{Part of the normalized self consistent quasiparticle spectrum
as obtained for  given values of
the normalized transverse energy $\varepsilon_\perp /E_F$.
The parameter $h_0/\Delta_0$ is chosen to be 0.8, and $k_F d_F=15$.
The region $|\epsilon_n/\Delta_0|>1$ is dominated by the
continuum sates, these are suppressed for clarity.
\label{spectra}}
 
\end{figure}

To provide additional perspective on the spectral
features of the proximity effect, we show
in Fig.~\ref{spectra} a relevant portion ($|\epsilon_n/\Delta_0|\leq1$)
of the calculated eigenvalues as obtained (under the same
conditions as in the previous Figure) for given
transverse
energies, $\varepsilon_\perp$. 
For a homogeneous superconductor, a gap separates
the continuum states that exist
for $|\epsilon_n/\Delta_0|>1$. 
With the inclusion of a ferromagnet, in addition
to the scattering states
there exists a discrete quasiparticle excitation spectrum
in the range $|\epsilon_n/\Delta_0|<1$ (see Fig.~\ref{spectra}).
The  bound states are visibly asymmetric 
about the
Fermi level due to the  effects the exchange field
has on the particle-hole amplitudes.
The existence of a finite number of bound-state branches
of localized quasiparticles is also
characteristic of antisymmetric domain walls.\cite{virtanen}
In Fig.~\ref{spectra} we can further see
that the  minimum in $|\epsilon_n|$, which 
corresponds to the energy gap seen in Fig.~\ref{gap} for
$k_F d_F = 15$, where $E_g\approx 0.09 \Delta_0$,
is  sharp, and that it occurs for quasiparticles whose
in-plane momentum is close to the Fermi level 
($\varepsilon_\perp/E_F \simeq 1$). This is
due in part to those quasiparticles not coupling 
to those states responsible for superconductivity.
This result is typical of what we find for other exchange fields, 
the only difference
being the value of $E_g$  [see Fig.~\ref{gap}].
It is therefore those excitations 
with a dominant momentum component 
parallel to the interface,
that are significant in contributing to the filling in of the gap.
In
the Andreev or quasiclassical approximation scheme, these 
quasiparticles have ``trajectories'' which are 
not treated accurately, and those with $\varepsilon_{\perp}/E_F>1$ are 
neglected
altogether.
This discrepancy has also been revealed\cite{geof} in
normal metal-superconductor
systems,
where  normal reflection  dominates the Andreev reflection
process for quasiparticles with large momentum parallel to the interface.

\begin{figure}
\includegraphics{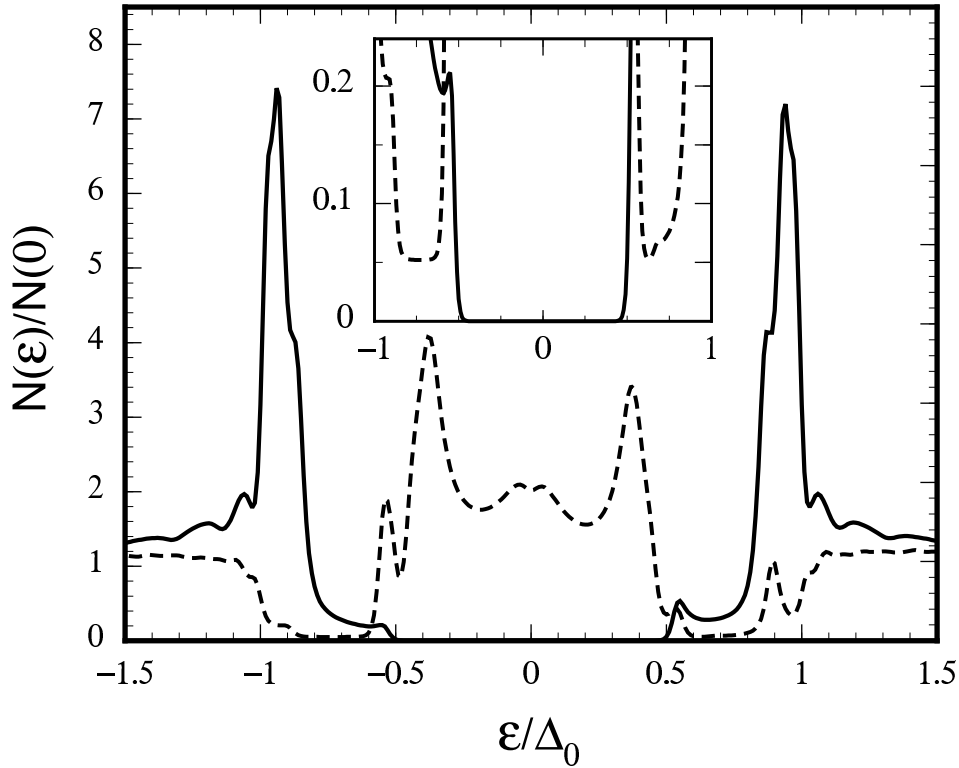}
\caption{Density of states in the ferromagnet (see text) for two exchange 
energies and a fixed width $k_F d_F = 9$. The bold curve corresponds
to  $h_0/\Delta_0=0.8$, and the dashed curve
to   $h_0/\Delta_0=8$. 
The DOS is normalized to $N(0)$, the DOS of the superconductor
in the normal state, and the energies are normalized to $\Delta_0$. 
The inset reveals the same quantities, magnified to clarify
the gap region. As in previous Figures, we assume
no interfacial scattering and no Fermi wavevector
mismatch: $Z=0$, $\Lambda=1$. \label{dos1}}
\end{figure}

The previous results in Fig.~\ref{gap} showed the strong influence
the exchange energy  has on $E_g$. 
Another quantity of great experimental interest is the local density of
one particle excitations in the magnet. 
Current experimental tools such
as the scanning tunneling microscope (STM) have atomic
scale resolution, and make this quantity experimentally accessible.
When  well defined
quasiparticles exist, the tunneling current is simply expressed 
as a convolution of the one-particle spectral function of the STM tip
with the spectral function for the F/S system.\cite{gygi}
The resultant tunneling conductance, which
is proportional to the density of states (DOS), is then
given as a sum of the individual contributions to the DOS 
from each spin channel.
\begin{equation}
\label{tdos}
N(z,\varepsilon)= N_\uparrow(z,\varepsilon)+N_\downarrow(z,\varepsilon), 
\end{equation}
where
the local DOS for each  
spin state is given by
\begin{eqnarray}\label{dos}
{N}_\uparrow(z,\epsilon) 
&=&-\sum_{n}
\Bigl\lbrace[u^\uparrow_n(z)]^2
 f'(\epsilon-\epsilon_n) 
+[v^\uparrow_n(z)]^2
 f'(\epsilon+\epsilon_n)\Bigr\rbrace, \\ 
{N}_\downarrow(z,\epsilon) 
&=&-\sum_{n}
\Bigl\lbrace[u^\downarrow_n(z)]^2
 f'(\epsilon-\epsilon_n) 
+[v^\downarrow_n(z)]^2
 f'(\epsilon+\epsilon_n)\Bigr\rbrace.
\end{eqnarray}
Here thermal broadening is accounted for
by the derivative of the
Fermi function, $f'(\epsilon) = \partial f/\partial \epsilon$.
To  investigate further how the previous results
interrelate with the local DOS (Eq.~\ref{tdos}), we show in 
Fig.~\ref{dos1} the 
local DOS $N(\epsilon)$, defined as $N(z,\varepsilon)$ integrated
over the  ferromagnetic layer. This
quantity is shown for two exchange fields, differing by a factor
of ten. The ferromagnet has a thickness of $d_F =9 k_F^{-1}$:
from Fig.~\ref{gap} this corresponds to an energy gap of 
$E_g\approx 0.5$ for $h_0/\Delta_0=0.8$,
and an absence of an energy gap for $h_0/\Delta_0=8$. This is consistent with
the DOS calculations, as Fig.~\ref{dos1} illustrates:
for the smaller exchange field, there is 
a complete absence of states for $\epsilon/\Delta_0 \lesssim 0.5$,
while the DOS is gapless at all energies for $h_0/\Delta_0=8$.
The observed energy gap for $h_0/\Delta_0=0.8$ is a result of the 
underlying competition 
between quantum size effects and constructive interference
between the spin-split quasiparticle amplitudes.
As expected, the particle-hole asymmetry is more prevalent in
the stronger magnet (within the subgap region), as
the two uppermost peaks
are clearly shifted relative to one another.

\begin{figure}
\includegraphics{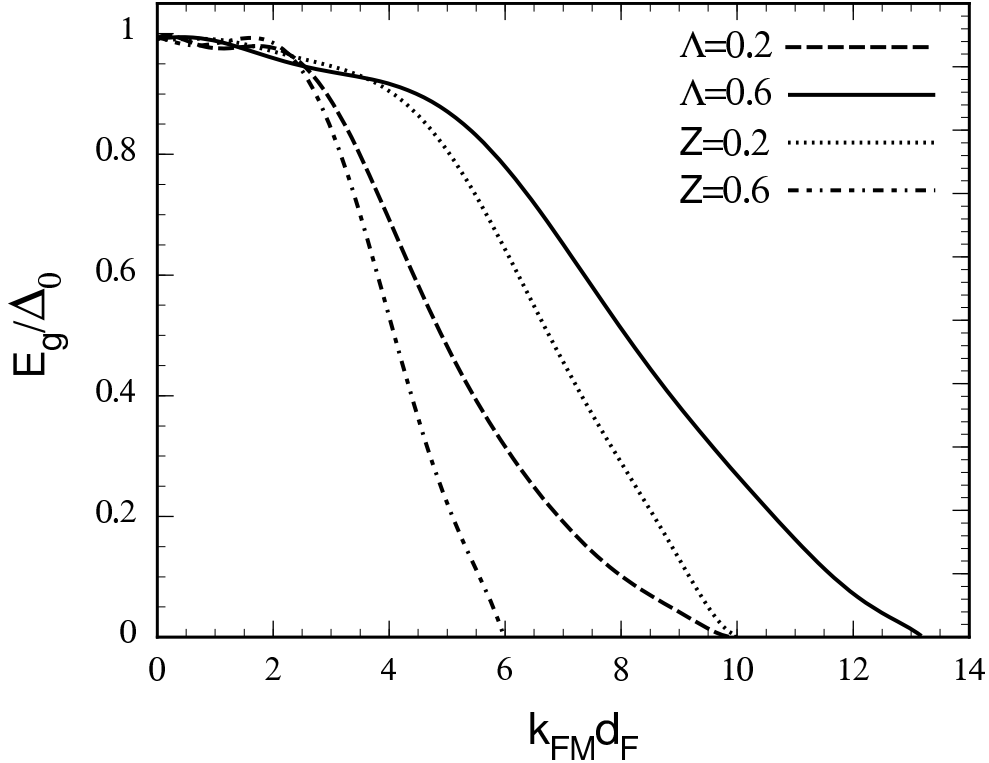}
\caption{Influence of interface scattering
and Fermi energy mismatch on the energy gap as a function of the
ferromagnet thickness. Both the effects of increasing the 
mismatch (decreasing $\Lambda$ from unity, see text) and 
increasing scattering strength 
(increasing $Z$) serve to
destroy the energy gap $E_g$ at an increased rate.\label{gap2}}
\end{figure}

\begin{figure}
\includegraphics{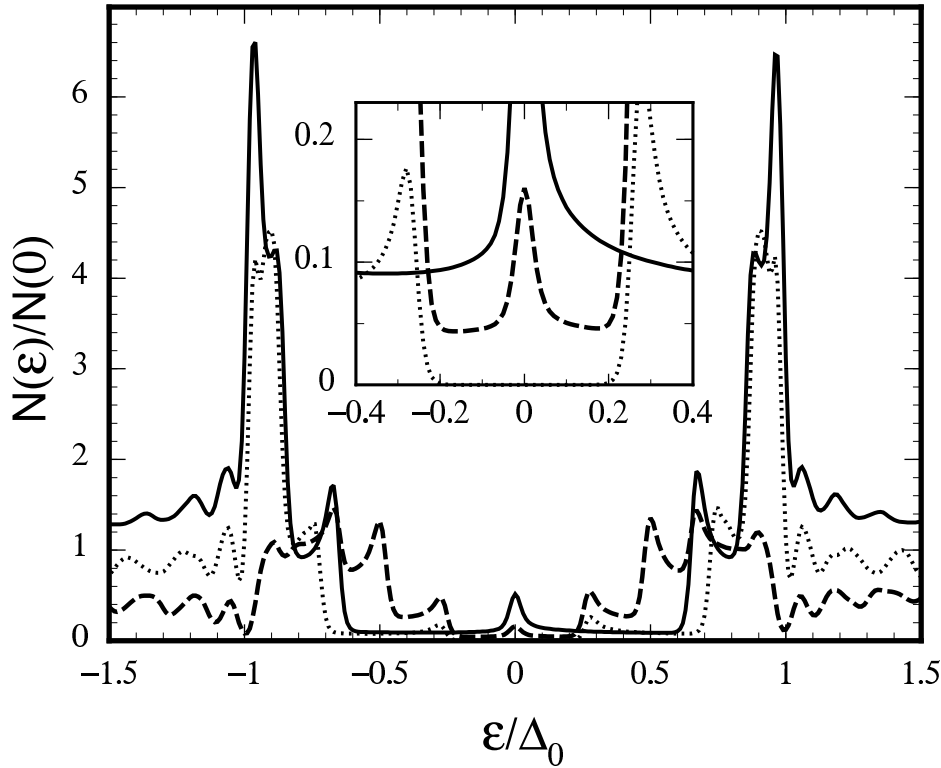}
 \caption{Local DOS in the 
ferromagnet of 
width $k_{FM} d_F=10$. The bold curve corresponds
to $Z=0.2$, and the dotted and dashed curves
are for the cases $\Lambda=0.6$, and $\Lambda=0.2$ respectively.
For the given dimensions of the magnet, the inset 
reveals that only
when $\Lambda=0.6$ does a gap survive in
the energy spectrum. At higher energies the curves
are clearly displaced from one another by a factor of
$\sqrt{\Lambda}$, due to the normalization. \label{dos2}}
 
\end{figure}

Next we consider the effect  that an interface barrier
or  Fermi energy mismatch has on the electronic structure. 
Experimentally, some degree of
Fermi energy mismatch is unavoidable, particularly with alloys,
and the transparency of the interface can be degraded due to
a thin oxide layer. 
In Fig.~\ref{gap2} we illustrate the normalized energy gap $E_g$ as a function
of $k_{F M} d_F$ (where $k_{FM}$ is the wavevector
corresponding to $E_{FM}$), for two  $\Lambda$ and $Z$ 
values that each differ by a factor of three. 
We fix the exchange field at
$h_0=0.8 \Delta_0$, keeping in mind that higher fields obey similar trends
so that generality is not lost. Also, in order to isolate each 
of the effects, whenever $\Lambda$ is
different from unity, $Z$ is zero, and vice versa. The figure shows that
increased mismatch (smaller $\Lambda$)
or increased interface scattering (larger $Z$),
induces the same trend
of reducing
$E_g$ as a function of $k_{FM}d_F$.
The functional form of the graphs are
distinct however:
for  $\Lambda <1$, the tail-end decay is slower
than the near linear decay in $E_g$
exhibited in the case of finite $Z$.
Further differences in the effects of the two parameters are
even more evident in the DOS spectra. To this end, Fig.~\ref{dos2}
shows the DOS in the ferromagnet for two different values of
the mismatch parameter $\Lambda$, and one value
of $Z$. Upon comparison with the previous Fig.~\ref{gap2},
we again find consistency between the DOS and
$E_g$ calculations:  the
energy gap is reduced at $k_{FM} d_F=10$ for
both $\Lambda=0.2$, and $Z=0.2$. This is reflected in
the DOS, where Fig.~\ref{dos2} illustrates a small but finite number of states
centered about the Fermi level.
Although the quasiparticle spectrum has just
turned gapless for $\Lambda=0.2$, and $Z=0.2$, the DOS
signature is substantially different, the most prominent feature
being the larger number of subgap bound states for $\Lambda=0.2$,
which can be attributed to the larger Fermi energy mismatch, and
subsequently, increased normal reflections.

\section{Conclusions \label{conclusions}}
We have self-consistently calculated the electronic structure  
for a F/S  heterostructure consisting of a 
thin ferromagnet layer adjoining a bulk superconductor. 
Our fully microscopic method
revealed the 
dependence of the transition to a gapless superconducting
state on the relevant physical parameters such as
exchange field energy, interface transparency, and Fermi energy mismatch.
The energy gap decreases with
the width of the F layer at a rate that depends upon the dimensionless
ferromagnet width $k_{FM} d_F$. The characteristic  length 
for gap suppression is much smaller
than the  length scale describing the decay of the
pair amplitude in the ferromagnet:
thus $E_g$ is suppressed more rapidly than
the pair amplitude.
Fermi level mismatch and interface barrier both serve to reduce the
gap further, however the functional forms of the respective reductions are
different. The local DOS in the ferromagnet is in all cases consistent
with the above observations.
The self-consistent quasiparticle spectra reveals that
the gap onset is due to
quasiparticle states with
a large momentum component parallel to the interface.
These states however,  occupy a relatively small
region of momentum space, and when 
calculating integral quantities over the full solid angle,
good agreement between our exact theory and quasiclassical theory
might be possible.

We have focused here on the 
spectral properties of the ferromagnet, in the clean limit.
This regime is most appropriate for layers whose dimensions do not exceed
the mean free path ($d_F \lesssim \ell$). Since, we have considered only 
thin layers, 
this assumption
is consistent with our calculations.
For situations where
the effects of impurities or
inelastic effects may be important,
a finite DOS at low energies would likely arise.
The general method used here however has found good overall
agreement with experiments involving
thin nonmagnetic normal metal-superconductor
bilayers, and bulk F/S structures.\cite{klaus}
Furthermore, previous calculations\cite{xing,zareyan2} within the 
ballistic limit have also
found agreement with experiment.
Other complexities may arise in actual experimental conditions, such
as grain boundaries and domain walls not accounted for here.
The fabrication of
relatively clean heterostructures,\cite{upad,soulen}
and structures with highly transparent interfaces\cite{gold} is 
possible however. Our calculations are therefore realistic, and the results 
should be reproducible
experimentally.

\appendix
\section{Numerical method}
We solve Eqns.~(\ref{bogo1},\ref{bogo2})
by expanding the quasiparticle
amplitudes in terms
of a finite subset of a set of orthonormal basis vectors,
$u^{\uparrow}_n(z)=\sum_{q} u^{\uparrow}_{n q}\phi_q(z)$, and
$v^{\downarrow}_n(z)=\sum_{q} v^{\downarrow}_{n q}
\phi_q(z)$.
We use the complete set of eigenfunctions
$\phi_q(z)=\langle z \vert q \rangle = \sqrt{{2}/{d}}\sin(k_q z)$,
where 
$k_q = {q/\pi}{d}$,
and $q$ is a positive integer.
The finite range of the pairing interaction $\omega_D$
permits the number $N$ of such basis vectors
to be cut off in the usual way.\cite{proximity}
Once this is done,
we arrive at the following $2N\times2N$ matrix eigensystem,
\begin{eqnarray} 
\label{nset1} 
\left[  
\begin{array}{cc} 
H^{+} & D \\ 
D & H^{-} 
\end{array} 
\right] 
\Psi_n 
= 
\epsilon_n 
\,\Psi_n, 
\end{eqnarray} 
where 
$\Psi_n^T =
(u^{\uparrow}_{n1},\ldots,u^{\uparrow}_{nN},v^{\downarrow}_{n1},
\ldots,v^{\downarrow}_{nN}).$
The matrix elements  $H^+_{q q'}$ 
connecting $\phi_q$ to $\phi_{q'}$
are 
constructed from the term found in brackets in Eq.(\ref{bogo1}),
\begin{eqnarray}
\label{basic}
H^{+}_{q q'}&=&\left\langle q \left\vert -\frac{1}{2m}\frac{\partial^2}
{\partial z^2} +
\varepsilon_{\perp}+W(z)-E_F(z) \right\vert q' \right\rangle\, \nonumber \label{basic1}\\ 
&=&\left[\frac{k^2_q}{2m} + \varepsilon_{\perp}\right]\delta_{q q'}+
\int_{0}^{d} dz\, \phi_q(z)W(z)\phi_{q'}(z)
 - E_{F \uparrow} \int_0^{d_F} dz\, \phi_q(z)\phi_{q'}(z) \nonumber \\
&-&
E_{F}  \int_{d_F}^{d} dz\, \phi_q(z)\phi_{q'}(z).
\end{eqnarray}
The expression for $H^-_{q q'}$ is calculated similarly
using  Eq.(\ref{bogo2}). The ``off-diagonal" matrix elements $D_{q q'}$
are given as,
\begin{eqnarray}
D_{q q'}&=&\left\langle \,q \left\vert \Delta(z) \right\vert q' \,\right\rangle
=\int_{d_F}^d dz \,\phi_q(z)\Delta(z)\phi_{q'}(z).
\end{eqnarray}
After performing the integrations, Eq.(\ref{basic1}) can be expressed
as
\begin{eqnarray}
H^{+}_{q q'}&=&\frac{2 W}{d} \sin(k_q d_F)\sin(k_{q'} d_F)-
\frac{E_{F\uparrow}}{d}\left[ \frac{\sin[(k_{q'}-k_q)d_F]}{(k_{q'}-k_q)}
- \frac{\sin[(k_{q'}+k_q)d_F]}{(k_{q'}+k_q)} \right] \nonumber \\
&-& \frac{E_{F}}{d}\left[ \frac{\sin[(k_{q'}+k_q)d_F]}{(k_{q'}+k_q)}
- \frac{\sin[(k_{q'}-k_q)d_F]}{(k_{q'}-k_q)} \right], \qquad q\neq q',  
\end{eqnarray}
while the diagonal matrix elements are written as,
\begin{eqnarray}
H^{+}_{q q'}&=&\left[\frac{k^2_q}{2m} + \varepsilon_{\perp}\right]+
\frac{W}{d}\Bigl[1-\cos(2 k_q d_F)\Bigr]-
\frac{ E_{F\uparrow}}{d}\left[ d_F-\frac{\sin(2k_q d_F)}{2 k_q}\right] \nonumber \\
&-& \frac{E_{F}}{d}\left[ d_S+\frac{\sin(2k_q d_F)}{2 k_q}\right], \qquad q= q'.
\end{eqnarray}
The self-consistency condition, Eq.(\ref{del2}) is now transformed into,
\begin{equation}
\label{selfcon}
\Delta(z) = \frac{\pi \lambda}{k_F d}
\sum_{p,p'}
{\sum_q}
\int{d\varepsilon_\perp}
\left[
u^\uparrow_{n p}v^\downarrow_{n p'}+
u^\downarrow_{n p}v^\uparrow_{n p'}\right]
\sin(k_p z)\sin(k_{p'} z)
\tanh(\epsilon_n/2T),
\end{equation}
where $\lambda=g(z) N(0)$, and $N(0)$ is the DOS for both spins of the superconductor
in the normal state. The quantum numbers $n$  encompass
the continuous transverse energy $\varepsilon_{\perp}$,
and  the quantized
longitudinal momentum index $q$.
The matrix eigensystem Eq.~(\ref{nset1}) and 
the self-consistency condition (\ref{selfcon})
are then solved numerically, using an iterative scheme developed and described
in earlier work\cite{proximity,klaus}.




\end{document}